%% file: ms.tex
\documentclass{emulateapj}
\usepackage{natbib}
\usepackage{amsmath,  amsthm, amssymb, cases}
\newcommand{\ts}{t_{\rm stop}}
\newcommand{\p}{\partial}

\newcommand{\cm}{\; {\rm cm}}

\newcommand{\vdr}{v_{\rm dr}}

\newcommand{\Sc}{\rm Sc}
\newcommand{\ag}{\alpha_g}

\newcommand{\kfgm}{k_{\rm f}}
\newcommand{\Kfgm}{\mathcal{K}_{\rm f}}
\newcommand{\lfgm}{\lambda_{\rm f}}
\newcommand{\gfgm}{\gamma_{\rm f}}

\slugcomment{Draft Modified \today}

\shorttitle{Dissipative GI and Particle Stirring}
\shortauthors{A.N.\ Youdin}

\begin{document}

\title{Planetesimal Formation without Thresholds. I: Dissipative Gravitational Instabilities and Particle Stirring by Turbulence}
\author{Andrew N. Youdin}
\affil{Princeton University Observatory, Princeton, NJ 08544}

\begin{abstract}
We analyze the gravitational collapse of solids subject to gas drag in a protoplanetary disk.  We also study the stirring of solids by turbulent fluctuations to determine the velocity dispersion and thickness of the midplane particle layer.  The usual thresholds for determining gravitational instability in disks, Toomre's criterion and/or the Roche density, do not apply.  Dissipation of angular momentum allows instability at longer wavelengths, lower densities, and higher velocity dispersions than without drag.   Small solids will slowly leak into axisymmetric rings since initial collapse occurs over many orbits.  Growth is fastest when particle stopping times are comparable to orbital times.  Our analysis of particle stirring by turbulence is consistent with previous results for tightly coupled particles, but is generalized to loose coupling where epicyclic motions contribute to random velocities.  A companion paper applies these results to turbulent protoplanetary disks.

\end{abstract}

\section{Introduction}

The formation of planetesimals, the first generation of solids bigger than a kilometer, remains a poorly undersood, and much debated, stage of planet formation.  The paradigm of collisional coagulation is very successful in explaining the  subsequent growth of planetesimals into planets.  In this stage surface gravity ensures collisional growth of the  largest bodies \citep{gls04}.  This bottom-up ``planetesimal hypothesis" \citep{lis93} naturally explains the remnant small bodies and impact craters so prevalent in our solar system.  Since small dust grains coagulate due to electrostatic interactions, it is tempting to interpolate that binary collisions are universally responsible for the growth of solids.  

However over a wide range of sizes, extending roughly from millimeter to kilometer radii, a robust sticking mechanism has not been identified, despite considerable experimental and theoretical effort (see \citealp{ys02}, hereafter YS02; \citealp{you04}; and references therein).   
Crossing this gulf of $\sim 10^{18}$ orders of magnitude in mass is particularly difficult because aerodynamic drag induces inward radial drift of solids with peak speeds of $50$ m/s for  $0.1 - 1$ meter solids \citep{stu77}.  This would result not only in destructive impacts, but also in  the loss of solids to the star in $\sim 100$ yr.  

Collective instabilities involving self-gravity and gas drag could directly assemble small solids (perhaps millimeter-sized to resemble known meteoritic inclusions) into planetesimals \citep{saf69, gw73}, thereby leapfrogging the regime where collisional binding energies are weak and collisional and drift velocities are fast.  The lone obstacle to gravitational instability of solids (hereafter GIS, or GI when referring to gravitational instability in general) is that turbulence stirs solids, lowering space densities and increasing velocity dispersions \citep{wc93}.  
Turbulence is thought to be prevalent in protoplanetary disks, mainly to drive stellar accretion \citep{sto00}.  Even in a passive disk, particle settling generates vertical shear in rotational velocities.  This can trigger turbulence and impede further particle settling  \citep{stu80}.

\citet{sek98} and YS02 showed that turbulence driven by vertical shear has a limited ability to loft solids.  In metal rich disks, with solid to gas ratios several times cosmic abundances, not all solids are stirred, resulting in midplane GIS.   The radial drift mentioned above can also serve as a particle enrichment mechanism.   Particles smaller than 0.1 meter migrate more slowly and ``pile-up" in the inner disk (YS02; \citealp{yc04}, hereafter YC04).  These works showed that rapid, dynamical GIS is possible.  The current work demonstrates that slower instabilities exist, and are easier to trigger. 

A common misconception is that GIS only occurs when the particle densities exceed the Roche limit or when Toomre's parameter, $Q_T$ (equation \ref{QT}), is less than unity.  For single component disks, whether fluid or particulate, these criteria are well-established \citep{too64, glb65I}, and for thin disks the two are usually equivalent (see \S\ref{nodiss}).  Similar criteria apply to a disk of stars and gas, where the coupling is gravitational \citep{raf01}.

A disk of solids and gas is coupled by dissipative drag forces as well.  If one assumes perfect coupling between solids and gas, no dissipation occurs and the standard Roche criterion for GI is recovered, within factors of order unity \citep{sek83, ys04}.  However when slippage is allowed, instability always occurs, even if the density is well below the Roche limit.   Dissipation of angular momentum ensures growth of sufficiently long wavelengths.  This result will be rederived, but is not new \citep{war76, war00}.  Growth rates are small for low densities, but can be much faster than disk lifetimes.  The mistaken belief in threshold criteria led to the view that exceedingly weak turbulence could prevent GIS.  Since GIS develops at lower densities, stronger turbulence can be tolerated.

This paper investigates the stability of a self-gravitating disk of solids (represented as a continuous fluid) subject to gas drag in \S\ref{sec:model}.   Section \ref{sec:c&h} estimates the particle velocity dispersion and scale height generated by turbulent fluctuations.  Concluding remarks in \S\ref{disc} include a summary of previous works on GIS in \S\ref{previous}.  A companion paper \citep[hereafter Paper II]{y05b} applies the results of this work to turbulent protoplanetary disk models.

\input{tab1.tex}

\section{Dissipative Gravitational Instabilities}\label{sec:model}
We investigate the gravitational stability of solids in the midplane of a protoplanetary disk as a fluid subject to drag against a steady gas background.  A two fluid description \citep[hereafter YG05]{yg05} would be more general, but our analysis isolates the collapse of solids through gas component that is gravitationally stable.  Instabilities involving the collapse of solids together with gas would be more rapid, but require higher densities and are not described here.

The stability properties of our model are described by three dimensionless parameters: $\tau_s$, $Q_T$, and $Q_R$.  Drag coupling is measured by
\begin{equation}\label{ts}
\tau_s \equiv \Omega \ts\, ,
\end{equation}
the particle stopping time, $\ts$, relative to the orbital time $1/\Omega$.  We will  refer to the  $\tau_s \ll 1$, $\tau_s \gg 1$, and $\tau_s \sim 1$ regimes as tight, loose, and  marginal coupling, respectively.  

Self-gravity is measured by two parameters:
\begin{eqnarray}
Q_T &\equiv& {c \Omega \over \pi G \Sigma}\, \label{QT},\\
Q_R &\equiv& {h \Omega^2 \over \pi G \Sigma} \approx {\Omega^2 \over \pi G \rho}\, \label{QR},
\end{eqnarray}
where $\Sigma$ and $\rho$ are, respectively, the surface and midplane space densities of solids, $c$ gives the particle velocity dispersion, and $h \sim \Sigma/\rho$ is the thickness of the particle subayer.  The  \citet{too64} parameter $Q_T$ pits the stabilizing effects of pressure and angular momentum against self-gravity, while $Q_R$ is the ratio of the Roche density to the particle density.  When $c \approx \Omega h$, which is often the case in astrophysical disks, $Q_T \approx Q_R$.  In \S\ref{sec:c&h} we show that a distinction is necessary when particles are stirred by turbulence.  For generality, this section keeps $\tau_s$, $Q_T$, and $Q_R$ independent.
  
\subsection{Basic Equations}
We model the evolution of solids in a localized patch of a protoplanetary disk.  At cylindrical radius $R = R_o$ we erect a Cartesian coordinate system with $x$ and $y$ axes directed in the radial and azimuthal directions, respectively.   The coordinates rotate uniformly at $\Omega_o = \Omega(R_o)$, the orbital frequency of the solids at $R_o$.  Our two dimensional model evolves the surface density and velocity of solids:
\begin{eqnarray}
\Sigma &=& \Sigma_o[1 + \sigma(x,y,t)]\,  \\
{\bf V} &=& {\bf V}_o(x) + {\bf v}(x,y,t) \, .
\end{eqnarray}
which are decomposed into steady background, $\Sigma_o$ and ${\bf V}_o(x)$, and fluctuating components, where $\sigma$ is a dimensioinless overdensity and ${\bf v} \equiv  u \hat{x} + v\hat{y}$.   Steady quantities vary on scales $\sim R_o$ (by assumption of smooth power-law profiles), and are spatially constant in the local approximation.  The exception is orbital velocity, given to first order in
$x/R_o$ by ${\bf V}_o\cdot\hat{y} = -q\Omega_o x $, with $q = -d\ln \Omega/d \ln R = 3/2$ for Keplerian shear.

Gas drag induces steady state radial drift of particles (inward for radially decreasing gas pressure).  In a global model, $\Sigma_o$ would be redistributed in a drift time (YS02, YC04), so our local approximation is strictly valid for shorter timescales, a condition that will be tested in Paper II.  Since radial motion is spatially constant in the local approximation, we remove it via a Galilean transformation and set ${\bf V}_o\cdot\hat{x} = 0$.  This is not possible in a two-fluid model with different drift speeds (YG05).

Perturbations evolve via height-integrated continuity and Euler equations with a drag acceleration.  To lowest order in $x/R_o$ these read:
\begin{eqnarray}
\frac{\p \sigma}{\p t} - q \Omega_o x \frac{\p \sigma}{\p y}  + \nabla\cdot[(1+\sigma ) {\bf v}] &=& 0\, , \label{cont}\\
\frac{\p {\bf v}}{\p t} -q\Omega_o x\frac{\p {\bf v}}{\p y} - q\Omega_o u \hat{y} + 2{\bf \Omega}_0 \times {\bf v} &=& -\nabla\chi - \frac{{\bf v}}{\ts } \label{mom}\, ,
\end{eqnarray}
Equations (\ref{cont}) and (\ref{mom}) are analogous to equations (34) and (37) in \citet{glb65II} where a detailed derivation can be found.  The linear drag acceleration, $-{\bf v}/\ts$, encompasses Epstein's and Stokes' laws \citep{ahn76} and applies to particles with radii $a \lesssim 10 (R/{\rm AU})^{5/4} \cm$.  We dropped small non-linear terms  $\mathcal{O}({\bf v}^2)$.  Our linear and axisymmetric analysis will also ignore the  $\mathcal{O}(\sigma{\bf v})$ term and set $\p /\p y = 0$.  The use of fluid equations is valid for processes involving timescales longer than $\ts$.  Thus loosely coupled solids can sometimes be treated as a fluid, for instance in calculating drift rates \citep{nsh86}.

The effective potential, $\chi \equiv \Phi + \Pi$, includes the perturbed gravitational potential, $\Phi$, and an effective pressure (more precisely, enthalpy):
\begin{equation}\label{EOS}
\Pi = c^2 \ln(1+\sigma)\, .
\end{equation}
The approximation that the velocity dispersion of solids acts as a thermodynamic pressure is a useful, if inexact, analogy that has been explored extensively in collisionless stellar dynamics \citep{bt}.  Its appropriateness for the current problem should be investigated in more detail.  Our choice of an isothermal equation of state is not crucial.  A different choice, e.g.\ $\Pi = c^2 (1+\sigma)^n$, would give similar results for linear perturbations.

The gravitational potential of an individual Fourier mode is:
\begin{equation}\label{poiss}
\tilde{\Phi} = -2 \pi G \Sigma_o {\tilde{\sigma} \over k} \mathcal{T}(kh)\, 
\end{equation}
where $\tilde{f}(k_x,k_y)$ is the 2D Fourier transform of $f(x,y)$ and $k \equiv \sqrt{k_x^2 + k_y^2}$.  The softening term, $\mathcal{T}(kh) = 1/(1+kh)$ mimics finite disk thickness in a vertically integrated model \citep{van70, shu84}.  This factor connects the thin disk limit for long waves, $kh \ll 1$, to the three dimensional result for $kh \gg 1$.

\subsection{Axisymmetric Dispersion Relation}\label{sec:DR}
We find linear, axisymmetric solutions to equations (\ref{cont} --- \ref{poiss}) by assigning a Fourier dependence, $\exp[\imath(k x - \omega t)]$, to perturbed quantities.  The resulting dispersion relation reads:
\begin{eqnarray}\label{dimDR}
\left(\omega + {\imath / \ts}\right)\left[ \Omega ^2 - 2 \pi G \Sigma k \mathcal{T}(k h) + k^2 c^2 \right. \\
\left. - \omega \left(\omega + {\imath / \ts}\right)  \right] - {i  \Omega^2 /\ts}=0\, , \nonumber
\end{eqnarray}
where above and henceforth we drop the naught subscripts on background quantities.  Equation (\ref{dimDR}) agrees with the dispersion relations derived in \citet[eq.\ A-22 ]{war76} and \citet[eq.\ 31]{war00}.  We introduce a dimensionless growth rate and wavenumber, 
\begin{eqnarray}
\gamma &\equiv& -\imath \omega/\Omega\, \label{eq:gamma}\\
\mathcal{K} 
&\equiv& k \lambda_G/(2 \pi) = {\pi G \Sigma k / \Omega^2} \, \label{eq:K},
\end{eqnarray}
where $\lambda_G$ is the characteristic gravitational wavelength, and express the dispersion relation as a real cubic:
\begin{equation}\label{realcubic}
(\gamma + \tau_s^{-1})[\gamma(\gamma + \tau_s^{-1}) + F] - \tau_s^{-1} = 0,
\end{equation}
where 
\begin{equation}\label{F}
F(\mathcal{K}, Q_T,Q_R) \equiv 1 - {2 \mathcal{K} \over 1 + \mathcal{K}Q_R} + Q_T^2 \mathcal{K}^2 \, .
\end{equation}
The function $F$ encapsulates all relevant physics (rotation, self-gravity, and velocity dispersion) except for dissipation.  Since $d\gamma/dF = (dF/d\gamma)^{-1} < 0$ for $\gamma > 0$, growth is faster for lower values of $F$, i.e.\ stronger self-gravity.  Note that the minimum of $F(\mathcal{K})$ is always $< 1$, a fact we will exploit shortly.

\subsection{Non-dissipative GI}\label{nodiss}
We begin by showing that our model reproduces standard non-dissipative  results.  As $\tau_s \rightarrow \infty$, equation (\ref{realcubic}) reduces to $\gamma^2 = -F$ plus a neutral mode with $\gamma = \omega = 0$.  Thus instability requires $F < 0$.  A more familiar, but identical, form for the density wave dispersion relation reads:
\begin{equation}\label{dr1}
\omega^2  = \Omega^2 - 2 \pi G \Sigma k \mathcal{T}(k h)  + k^2 c^2\, .
\end{equation}
In the thin disk limit $\mathcal{T} \rightarrow 1$, or equivalently $Q_R \rightarrow 0$, we recover the familiar instability criterion $Q_T < Q_{T,\rm crit} = 1$, and the marginally stable wavenumber is $\mathcal{K}_{\rm crit} = 1$.

It is worth considering  finite thickness effects in the dissipation-free case.   In a cold disk with $Q_T = 0$, stability is assured when $Q_R \ge 2$.  The Roche limit is enforced \emph{in the absence of dissipation}.
For less extreme cases, as $Q_R$ increases (from $0\rightarrow 2$),  $Q_{T, \rm crit}$ decreases ($1\rightarrow 0$), and $\mathcal{K}_{\rm crit}$ increases ($1 \rightarrow \infty$).  Physically, at lower volume densities the disk must be colder to go unstable, and it will do so at shorter wavelengths.  If we require $Q_R = Q_T$ (the standard  $c = \Omega h$ relation) then  $Q_{T, \rm crit} \approx 0.55$ and $\mathcal{K}_{\rm crit} \approx 1.2$.  This actually overestimates the stabilizing influence of finite thickness.  Self-gravity shrinks the thickness of an isothermal fluid so that $Q_T = Q_R \sqrt{1+2/Q_R}$ (\citealp{war00}; our eq.\ [\ref{psi}]) includes the same correction for vertical self-gravity).  This gives $Q_{T, \rm crit} \approx 0.77$ and $\mathcal{K}_{\rm crit} \approx 1.05$, a less dramatic finite thickness correction.

\begin{figure}[htb]
 \hspace{-.4in}
 \includegraphics[width=3.9in]{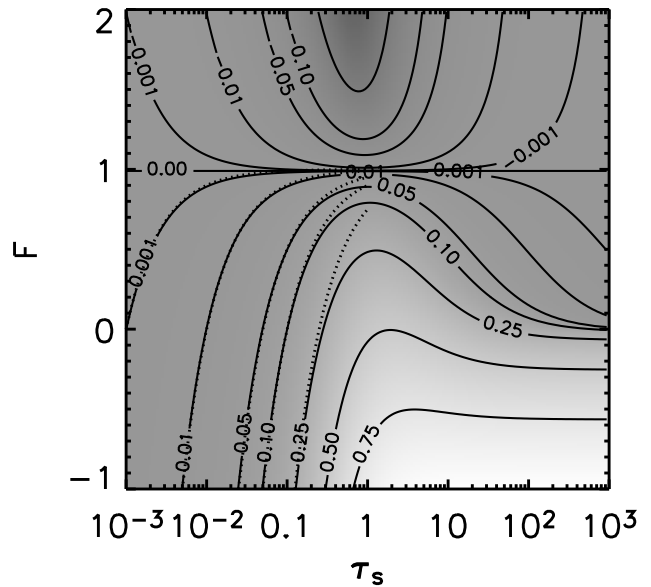} 
   \caption{Growth rate contours (normalized to orbital frequency) for dissipative GI as a function of stopping time and $F$ (eq.\ (\ref{F})].    The tight-coupling approximation is overplotted with dotted contours. See \S\ref{Flt1}.}
   \label{fig:cubgrow}
\end{figure} 

\subsection{Dissipative GI}
\subsubsection{Dissipation Guarantees Instability}\label{Flt1}
Drag forces damp angular momentum that would otherwise support long wavelength modes.  This qualitatively changes stability characteristics.  While $F<0$ was needed for instability in the absence of dissipation, equation (\ref{realcubic}) with $\tau_s$ real and positive implies growth for $F < 1$.  Appendix \ref{sec:proof} proves this necessary and sufficient condition.\footnote{Appendix \ref{sec:proof} also shows that only one of the three roots can grow, and that this mode does not oscillate.}  Since $F < 1$ as $\mathcal{K} \rightarrow 0$ (eq. [\ref{F}]), sufficiently long wavelengths are always unstable.  

Figure \ref{fig:cubgrow} shows growth rate contours as a function of $F$ and $\tau_s$.  The stability boundary at $F = 1$ is clear.  At large $\tau_s$, contours of slow but finite growth ($\gamma \lesssim 0.1$)  converge to $F = 0$, the dissipation-free result.  Growth rates  peak around $\tau_s = 1$ because dissipation is strong enough to damp vorticity on an orbital time, but not so strong that it impedes collapse.

\subsubsection{Fastest Growing Wavenumbers}\label{fgm}
The wavenumber giving the fastest linear growth rate, $\kfgm$ (or the dimensionless $\Kfgm $), is the most relevant until non-linear effects become significant.  Since all $\mathcal{K}$ dependence arises through $F$, $\Kfgm $ is the root of:
\begin{equation}
{d \gamma \over d \mathcal{K}} = {d \gamma \over d F}{d F \over d \mathcal{K}} = 0\, .
\end{equation}
Since $d \gamma/dF < 0$ for all growing modes, $\Kfgm $ is just the root of $d F / d \mathcal{K} = 0$ which satisfies:
\begin{equation}\label{kfgm}
\Kfgm Q_T^2(1+ \Kfgm Q_R)^2 = 1.
\end{equation}
Thus $\Kfgm$ is determined by the stability parameters, seemingly independent of dissipation.  Without dissipation the longer waves would not grow.   Also drag coupling influences $Q_T$ and $Q_R$ values (\S\ref{sec:c&h}).

Modes will be long (or short) compared to the disk scale height if $\kfgm h = \Kfgm Q_R \ll 1$ (or $ \gg 1$).  We refer to these as the thin (or thick) disk limits, respectively.  Note that $\kfgm h$ depends on $\Theta \equiv Q_T^2/Q_R$ only.  Series solutions show: 
\begin{subnumcases}{\label{kser} \kfgm h \simeq }
{1 \over \Theta} - {2 \over \Theta^2} 
+ \mathcal{O}(\Theta^{-3}) & if $\Theta \gg 1$, \label{klong}\\
\label{kshort}  {1 \over \Theta^{1/3}} - {2 \over 3} + 
\mathcal{O}(\Theta^{1/3})\,& if $\Theta \ll 1$.
\end{subnumcases}
The $\Theta^{-1/3}$ dependence in equation (\ref{kshort}) means that extreme conditions (very cold and low density) are needed for waves to be much shorter than $h$.

The fastest growing wavelengths, $\lfgm = 2 \pi/\kfgm = \lambda_G/\Kfgm$, satisfy:
\begin{subnumcases}{\label{lfgm} {\lfgm \over \lambda_G} \simeq }
Q_T^2 & if $\kfgm h \ll 1$, \label{llong}\\
\label{lshort}  (Q_T Q_R)^{2/3} & if $\kfgm h \gg 1$.
\end{subnumcases}
 The thin disk limit is independent of scale height as it must be.  In both regimes, hotter disks ($Q_T$ large) disks go unstable at longer wavelengths to avoid ``pressure" stabilization.  Most importantly, modes will be much larger than $\lambda_G$ when the stability parameters are large.

\subsubsection{Tight Coupling}\label{tight}

Particle sizes and gas densities relevant to planetesimal formation usually correspond to tight coupling, $\tau_s \ll 1$.  In this regime a destabilized neutral mode grows at a rate
\begin{equation}\label{drtight}
\gamma \simeq \tau_s(1-F) + \mathcal{O}(\tau_s^3)\, .
\end{equation}
The two other modes are less interesting: strongly damped inertial oscillations with $\gamma \approx -\tau_s^{-1} \pm \imath$.  The dotted contours in Figure \ref{fig:cubgrow} show that equation (\ref{drtight}) agrees with the full solutions for small $\tau_s$.  Leading order solutions for $\Kfgm $ (equations \ref{lfgm}) in equations (\ref{F}, \ref{drtight}) give the fastest growth rates as:
\begin{subnumcases}{\gfgm  \approx}
\label{growlong} {\tau_s \over Q_T^2} &if $\kfgm h \ll 1$ ,\\
\label{growshort} {2 \tau_s \over Q_R}
& if $\kfgm h \gg 1$.
\end{subnumcases}
Growth in the thin disk limit depends on the velocity dispersion but not the scale height, and vice-versa in the thick disk limit.  For large values of the stability parameters, dissipative growth is slower than the particle settling rate, $\Omega \tau_s$.  Thus linear growth proceeds through a series of quasi-equilbrium states.  The loose coupling case is evaluated in Appendix \ref{weak}.

\subsection{Why Dissipation Helps}
To explain how angular momentum dissipation aids instability, we offer the following analogy with the Rayleigh instability.  The arguments are fairly technical and can be skipped if desired.  Generating axisymmetric density enhancements requires radial velocity perturbations, $u$.  In the absence of dissipation, the azimuthal force equation conserves angular momentum:
\begin{equation}\label{am}
{\partial v /\p t} = (-2+q)\Omega u
\end{equation}
or $ v = -(2-q)u/\gamma$, in terms of the dimensionless growth rate.  The  Coriolis force is partly blunted by differential rotation, with $q = 3/2$ in a Kepler disk.    For slow growth ($\gamma \ll 1$), angular momentum conservation requires $|v| \gg |u|$ unless $q\approx 2$.  Azimuthal motions in turn supply a radial Coriolis force, $f_{C,r} = 2 \Omega v = -2(2-q)u\Omega/\gamma$.  For $q<2$, $f_{C,r}$ decelerates $u$ and has the form of a drag force that becomes stronger for slower growth.  For $q>2$, the sign of the radial acceleration changes, giving rise to the well-known Rayleigh instability for an outwardly decreasing angular momentum gradient.

Now we include actual drag forces and restrict ourselves to $q = 3/2$.  Azimuthal force balance now includes a loss term for orbital angular momentum.
\begin{equation}\label{amdiss}
{\partial v /\p t} = -\Omega u/2 - v/\ts .
\end{equation}
If $\tau_s \gamma \ll 1$, i.e.\ strong drag or slow growth, the  Coriolis force (blunted by differential rotation) is balanced by drag.  The resulting azimuthal perturbation, $v \approx - \tau_s u/2$, is much smaller in magnitude than the drag free case (where $v \sim -u/\gamma$) as if $q$ were approaching 2.  In turn $f_{C,r}' = 2 \Omega v = - \Omega  \tau_s u$ (the prime differentiates between this and the drag-free case) offers much less resistance to collapse.  For $\tau_s \ll 1$, the radial component of the drag force, $f_{d,r} = -u/ts$, is the dominant obstacle to collapse.   Since $|f_{d,r}| \ll |f_{C,r}| \sim |u| \Omega/\gamma$ for $\gamma \ll \tau_s$, the presence of drag allows slow growth that would otherwise be prevented by angular momentum conservation.  Since this collapse is primarily resisted by drag, it strongly resembles particle settling (to the center of the contracting annulus) at the terminal velocity.

\section{Particle Stirring by Turbulence}\label{sec:c&h}
The previous section treats $Q_T$ and $Q_R$ as free parameters.    To constrain these stability parameters, this section calculates the velocity dispersion, $c$, and scale height, $h$, of solids coupled to turbulent fluctuations.  Our approximate expressions reproduce known results in the tight coupling limit, but we generalize the analysis to arbitrary stopping times.  In contrast to most astrophysical disks, we find $c \neq \Omega h$, with $c \ll \Omega h $ for weak turbulence and tight coupling.  This motivated our use of two stability parameters.   We ignore other contributions to random motion, like gravitational scattering, aerodynamic drift, or phyical collisions, which are less significant, particularly for small, tightly coupled solids, see \S\ref{sec:dc}.  

\subsection{Turbulent Diffusion}
We assume that gaseous turbulence acts a diffusive viscosity
\begin{equation}\label{eq:nu}
\nu_g \equiv \ag c_g^2/\Omega  \approx v_0^2 t_0 \, ,
\end{equation}
defined in terms of  $\ag$, a dimensionless transport coefficient for gas, and  $c_g$, the sound speed.  The second equality relates $\nu_g$ to the characteristic eddy speed, $v_0$, and turnover time, $t_0$.  We must specify two of $\ag$, $v_0$, and/or $t_0$ to describe the turbulence.

In general, particles are not perfectly coupled to turbulent fluctuations.  \citet{miz88} calculated the RMS velocity of solids, $c_\nu$, in response to a Kolmogorov spectrum of eddies.  They found $c_\nu \approx v_0$ for particles well coupled to the largest eddies ($\ts < t_0$), and $c_\nu \approx v_0 \sqrt{t_0/\ts}$ if  $\ts > t_0$.  Both regimes are covered by:
\begin{equation}\label{miz}
 c_\nu \approx {v_0 \over  \sqrt{1 + \tau_s/\tau_0}} = \sqrt{\ag \over \tau_0 + \tau_s}c_g \, .
 \end{equation}
where $\tau_0 \equiv \Omega t_0$.

The diffusivity of the solids, $\nu_s \equiv \alpha_s c_g^2/\Omega$, is defined analogously to $\nu_g$.  Assuming the velocity perturbations (equation \ref{miz}) fluctuate on the forcing time, $t_0$, we have $\nu_s \approx c_\nu^2 t_0$.  The Schmidt number, a ratio of particle to gas diffusivities, is then \citep{cdc93}:
\begin{equation}\label{Sc}
{\rm Sc} \equiv \nu_g /\nu_s = \ag/\alpha_s \approx 1 + \ts/t_0\, .
\end{equation}
As $\ts \rightarrow 0$, solids diffuse as efficiently as gas in this approximation.  

\subsection{Particle Scale Height}\label{sec:height}
We estimate an equilibrium scale height for the solids by equating diffusion times, $t_{\rm diff} = h^2/\nu_s$ with settling times, $t_{\rm sett}$.  Tightly coupled particles fall to the midplane at terminal velocity and $t_{\rm sett} = (\psi \Omega^2 \ts)^{-1}$, where self-gravity amplifies vertical gravity by a factor
\begin{equation}\label{psi}
\psi = 1 + 2/Q_R\, .
\end{equation}
For weak GI with $Q_R \gg 1$, we will frequently set $\psi = 1$. Loosely coupled particles undergo damped vertical oscillations with $t_{\rm sett} = \ts$, with no correction for self-gravity.   The combined result for arbitrary $\ts$ is:
\begin{equation}
t_{\rm sett} = {\ts + 1/(\psi \Omega^2 \ts)}\, .
\end{equation}

Equating $t_{\rm diff} = t_{\rm sett}$ and using equation (\ref{Sc}) gives $h$, here expressed relative to the gas scale height, $h_g = c_g/\Omega$:
\begin{equation}\label{hgen}
{h \over h_g} = \sqrt{{\ag}{1 + \psi(h)\tau_s^2 \over  \psi(h)\tau_s(1 + \tau_s/\tau_0)}}\, .
\end{equation}
Our diffusion time implicitly assumed that $h < h_g$.  If equation (\ref{hgen}) gives $h > h_g$, then  particles are evenly mixed with gas and $h \approx h_g$.  The $h$ dependence of $\psi$ is noted, because when self-gravity is included, equation (\ref{hgen}) defines $h$ implicitly as the root of a cubic.  In all cases, $h$ increases with $\ag$.

As $\tau_s \rightarrow \infty$, $h \rightarrow h_g \sqrt{\ag  \tau_0} = v_0 t_0$, 
a constant equal to the characteristic eddy size, $l_0$.    One might naively expect $h$ to decrease at large $\tau_s$, but in our analysis both the settling time and the diffusion time increase $\propto \tau_s$.  A non-linear contribution to $\Sc$ would drive $h\rightarrow 0$ in the loose coupling limit.  Since $h \sim l_0$ for $\tau_s \gg 1$,  the approximation that turbulence acts diffusively is only marginally justified in the loose coupling regime.  For tight coupling, $ h \gg l_0$, and the diffusion approximation is well justified.  

\subsection{Velocity Dispersion}\label{sec:vd}
The random motions of the solids includes direct kicks from turbulent fluctuations, $c_\nu$ (equation \ref{miz}), plus a contribution from epicyclic oscillations, $c_{\rm orb}$.  We assume that eccentricities, $e$, and inclinations, $i$, are small and in the dispersion dominated regime, $e \approx i\ll 1$.  The net velocity dispersion,
\begin{equation}\label{ctot}
c = \sqrt{c_\nu^2 +c_{\rm orb}^2}\, 
\end{equation}
is dominated by $c_\nu$ for tight coupling, and $c_{\rm orb}$ for loose coupling.

To estimate $c_{\rm orb}$, consider epicyclic oscillations of an individual particle with gas drag :
\begin{equation}\label{epi}
\ddot{x} = -\Omega^2x - \dot{x}/ts\, ,
\end{equation}
We isolate radial motions since we are concerned with support against radial collapse (see \citealp{gbl04} for a detailed treatment of vertical oscillations).
The epicyclic frequency  needs no correction for self-gravity since the total disk mass is small compared to the central star.  Equation (\ref{epi}) describes a damped harmonic oscillator.  Driving by turbulent fluctuations is not included, but should maintain a constant amplitude $h \approx i R \approx e R$.\footnote{Ideally turbulent driving and orbital dynamics would be treated concurrently, a task beyond the scope of the current work.} 

We define $c_{\rm orb} = v_{\rm max}$, where $v_{\rm max}$ is the maximum speed achieved by a particle released at rest from $x = h$.  For $\tau_s \gg 1$, $v_{\rm max} \approx \Omega h$ as the particle crosses $x = 0$, the radius of the orbit's guiding center.  With $\tau_s \ll 1$ the particle reaches terminal speed $v_{\rm max} \approx \Omega \tau_s h$ shortly after release.  Intermediate cases reach a maximum speed:
\begin{equation}\label{corbgen}
c_{\rm orb} = {\Omega h \tau_s \over 1 + \tau_s} = \sqrt{\ag \tau_0 \over  \psi}\sqrt{\tau_s(1 + \psi \tau_s^2) \over \tau_0 + \tau_s}{c_g \over 1 + \tau_s} \, ,
\end{equation}
using equation (\ref{hgen}) to eliminate $h$.

Comparison of equations (\ref{miz}) and (\ref{corbgen}) shows that $c_{\rm orb}$ dominates random motions for $\tau_s \gtrsim 1/\tau_0$, while $c_\nu$ is larger for tighter coupling.  The dynamically interesting ratio
\begin{equation}
{c \over \Omega h} = {Q_T \over Q_R} = {\tau_s \over 1 + \tau_s}\sqrt{1+ {\psi (1 + \tau_s)^2 \over \tau_0 \tau_s(1 + \psi \tau_s^2)}}.
\end{equation}
If particles are well coupled to eddies, $\tau_s < \tau_0$, and to orbital motions, $\tau_s \ll 1$, then $c/(\Omega h) \approx \sqrt{\psi \tau_s/\tau_0} \ll 1$.  In the loose coupling limit we recover the standard $c/(\Omega h) \approx 1$.  For eddies with short turnover times, $\tau_0 \ll 1$, $c/(\Omega h) \approx \sqrt{2/\tau_0} \gg 1$ at $\tau_s = 1$, showing that a wide variety of behaviors are possible.
\begin{figure}[tb] 
  \hspace{-.3in}
   \includegraphics[width=3.8in]{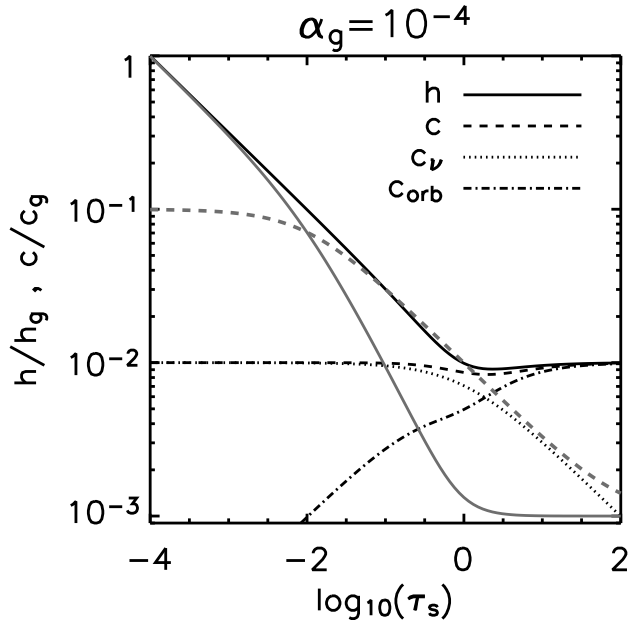} 
\vspace{-.2cm} 
   \caption{Particle scale height, $h$ (\emph{solid curves}), and velocity dispersion, $c$ (\emph{dashed curves}), vs.\ stopping time for turbulence with orbital turnover times (\emph{black curves}) and faster eddies (\emph{grey curves}) with $\ag = 10^{-4}$ in both cases.    For  orbital turnover times, the two contributions to $c$ are shown separately.  With faster eddies, $c_\nu$ dominates and only the total $c$ is plotted for clarity.} \label{fig:ch}
   \vspace{.2in}
\end{figure}

\subsection{Results: Slow vs.\ Fast Eddies}\label{totime}
Turbulence cannot be described by a single free parameter, $\ag$.  We must also specify the speed of eddies.  We consider two cases, though intermediate speeds are possible. 
\begin{figure}[tbp] 
  \hspace{-.3in}
  \includegraphics[width=3.8in]{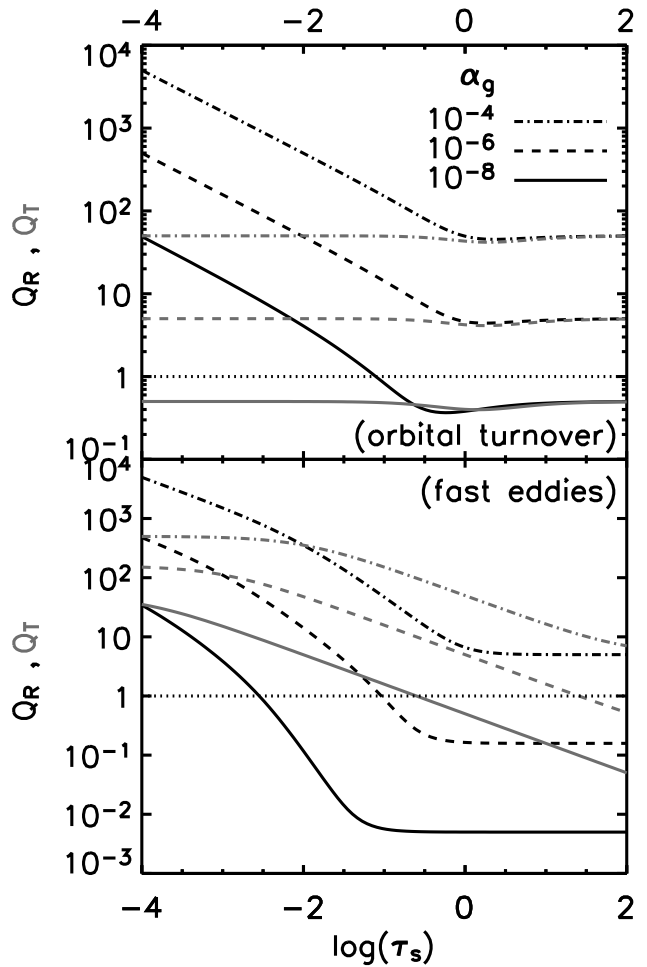} 
 \vspace{-.2cm} 
   \caption{\emph{Top:} Stability parameters $Q_R$ (\emph{black curves}) and $Q_T$ (\emph{grey curves}) vs.\ stopping time for orbital turnover times with $Q_{\rm sonic} = 5\times 10^3$.  The gas diffusivity, $\ag$, takes values of $10^{-8}$ (\emph{solid curves}), $10^{-6}$ (\emph{dashed}), and $10^{-4}$ (\emph{dot-dashed}). The dotted line orients the reference value of 1.  \emph{Bottom:} same as above but for faster eddies.}
   \label{fig:Qs}
   \vspace{.2in}
\end{figure}
\subsubsection{Orbital Turnover Times} \label{orbturb}
If orbital shear is involved in the generation or destruction of turbulence one naturally expects $\tau_0 = 1$.  Since eddies with  longer turnover times are destroyed by radial shear, this gives the slowest allowed eddy velocity,  $v_0 = \sqrt{\ag}c_g$.  Weak turbulence with $\alpha_s \ll 1$ is very subsonic.  Values for $h$ and $c$, plotted in Figure \ref{fig:ch} with black curves, are obtained from equations (\ref{hgen}), (\ref{miz}), and (\ref{corbgen}) with $\tau_0 = 1$ and $\psi = 1$.  When $\ag < \tau_s < 1$, coupling is imperfect enough to allow some settling and we recover the standard result (\citealp{dms95}, also in YC04):
\begin{equation}\label{hkep}
h \approx h_g \sqrt{\ag / \tau_s}\, .
\end{equation}
For loose coupling $h \approx \sqrt{\ag} h_g$.  The velocity dispersion maintains a nearly constant value $c \approx \sqrt{\ag} c_g$, but the source shifts from direct kicks by turbulent fluctuations for $\tau_s \ll 1$ to  epicyclic motion for $\tau_s \gg 1$.  For tight coupling $c/(\Omega h) \approx \sqrt{\tau_s} \ll 1$ because small velocity kicks loft particles efficiently.  

Expressions for the stability parameters are simplified by defining 
\begin{equation}\label{Qsonic}
Q_\mathrm{sonic} \equiv {c_g \Omega \over \pi G \Sigma},
\end{equation}
a ``mixed" Toomre parameter involving the gas sound speed and particle surface density.The stability parameters are approximated by
\begin{subequations}\label{Qorb}
\begin{eqnarray}
Q_R &\approx& 
\begin{cases}
Q_{\rm sonic} &\textrm{if $\tau_s < \ag$} \\
\sqrt{\ag/\tau_s} Q_{\rm sonic} &\textrm{if $\ag<\tau_s < 1$} \\
\sqrt{\ag} Q_{\rm sonic} &\textrm{if $\tau_s > 1$} \\ 
\end{cases} \\
Q_T & \approx& \sqrt{\ag} Q_{\rm sonic} \, .
\end{eqnarray}
\end{subequations}
Numerical solutions are plotted for several values of $\ag$ in Figure \ref{fig:Qs} (\emph{top}).  Stability parameters are larger, i.e.\ self-gravity is weaker, for stronger turbulence.  For tight coupling we confirm  $Q_T/Q_R = c/(\Omega h) \approx \sqrt{\tau_s} \ll 1$.

\subsubsection{Fast Eddies}\label{sonicturb}
We now consider what happens if turnover times are shorter and eddy speeds are faster.  We still want eddies to be subsonic for $\ag \ll 1$, so we choose $v_0 = \ag^{1/4} c_g$.  In this case turnover times are short, $\tau_0 = \sqrt{\ag} \ll 1$, and eddies are small, $l_0 \sim \ag^{3/4} h_g$.  For comparison recall that with orbital turnover times we had $\tau_0 = 1$, $v_0 = \sqrt{\ag} c_g$, and $l_0 =  \sqrt{\ag} c_g$.  We favor the case of orbital turnover times, and the case of fast eddies (FE) is primarily meant to be instructive.\footnote{We also investigated the extreme case of $v_0 = c_g$, and the results were qualitatively similar to FE.  Since sonic turbulence implies shocks, our formalism might not be appropriate, and we do not present those results.}

Figure \ref{fig:ch} (grey lines) plots $c$ and $h$ for FE.  Compared to turbulence with orbital turnover times (black lines) $h$ is smaller, and thus densities higher, because particles are less tightly coupled to eddies once $\tau_s > \tau_0 = \sqrt{\ag}$.  For $\tau_s < 1$ FE generate a larger velocity dispersion.  We know from \S\ref{tight} that, in the long wave limit, larger $c$ gives slower growth even if $h$ is smaller.  Paper II confirms that this is the correct regime to consider.  Thus growth is slower for tightly coupled particles for FE.\footnote{This conclusion may not hold when collisional dissipation, which Appendix \ref{sec:dc} shows is significant for $c \gg \Omega h$, is included.}  For $\tau_s > 1$ however, particles are so poorly coupled to FE that $c$ is lower despite the higher eddy speed.  Thus for loosely coupled solids GIS is more rapid  for FE (with $\ag$ fixed).   

Figure \ref{fig:Qs} (\emph{bottom}) plots the stability parameters for the fast eddy case.  For weak turbulence and loose coupling, values of $Q_R \approx \ag^{3/4} Q_{\rm sonic}$ are remarkably small.  However, as long as $Q_T \gtrsim 1$ violent dynamical instability is prevented.  Both plots (Fig. \ref{fig:ch} and Fig. \ref{fig:Qs}) show $c \gg (\Omega h)$ for fast eddies in the range $\sqrt{\ag} < \tau_s < 1/\sqrt{\ag}$.  Thus FE produce a thin, hot disk.

\section{Discussion}\label{disc}
\subsection{Previous Work}\label{previous}
Many authors have considered GI of solids in a gas disk.  \citet{saf69} and \citet{gw73} are most associated with planetesimal formation by GI.  Neither included dissipation by drag in their stability criteria.    Instead they used the Roche and Toomre criteria, respectively.  Both of these give GI at the characteristic  wavelength $\lambda_G$.  In minimum mass models this leads to the kilometer estimate for planetesimal sizes.  The use of threshold criteria led later researchers to conclude that turbulence could readily prevent planetesimal formation by GI \citep{stu80}.   \citet{gw73} did use drag to explain dissipation of angular momentum in the second stage of their two-stage collapse model.  Moreover, they noted that ``gas drag does destabilize axisymmetric perturbations for wavelengths larger than ($\lambda_G$)," but did not investigate the possibility in detail.

Subsequent work has  included dissipation to find GI at low densities, but the implications  have not been widely appreciated.
\citet{spi72} considered the Jeans instability, i.e.\ gravitational collapse in a uniform non-rotating medium, in a coupled two fluid system.  Spiegel described ``clumping of dust alone" as it ``slips through the gas."  \citet{war76, war00} considered the effect of a linear friction term on GI in disks.  He found ``destabilized neutral modes" always exist at long wavelengths.  Our \S\ref{sec:model} is a more detailed investigation of the same basic model.  Ward analyzed the dispersion relation in the long-wavelength limit, which allowed him to avoid estimating the velocity dispersion or scale height.  Since we isolate the fastest growing mode, our growth rates are somewhat faster. 

\citet{cor81} investigated GI in a disk of two coupled fluids, but the results are confusing.  A minimum density threshold for GI is claimed, even though they derive a growth time that reproduces the \citet{war76} result with no density cutoff.  The problem appears to be that  \citet{cor81}  arbitrarily impose a maximum wavelength cutoff by appealing to the ``conservation of angular momentum of the grains" even while  correctly noting that ``angular momentum can be freely exchanged with the stationary gas."

\citet{gp00} primarily considered drag instabilities.  They also found long wavelength GI with slow growth rates for certain choices of nebular parameters.  Since they use a different drag prescription (boundary layer or ``plate" drag) and include diffusive effects, direct comparison with our work is difficult.

The response of solids to a gravitationally unstable gas disk has been considered.  \citet{nvc91} found that adding solids increased the growth rate of global, non-axisymmetric GI.  \citet{ric04} showed that 1-10 meter planetesimals accumulate in the spiral arms of a massive gas disk.  The literature also contains many mechanisms for concentrating solids without self-gravity (see YS02 for an incomplete summary).

\subsection{Summary and Future Work}
This work investigated the gravitational collapse of solids subject to gas drag.  We explored the dependence on layer thickness and velocity dispersion (normalized to the strength of self-gravity by $Q_R$ and $Q_T$, respectively), and the aerodynamic stopping time, $\tau_s$.  Even when self-gravity is weak, sufficiently long wavelengths are always unstable to collapse over many orbital times.  This initially forms a large circumstellar annulus which should subsequently fragment to form less massive planetesimals.

We also investigated particle stirring by turbulence, which should largely determine the velocity dispersion, $c$, and scale height, $h$.  In contrast to the usual $c \approx \Omega h$ relation, we found $c \ll \Omega h$ for tightly coupled particles when eddy turnover times are orbital.  Paper II uses these results to investigate the gravitational accumulation of solids in turbulent disks.

Our dynamical model could be refined by adding physics such as  a dispersion of particle sizes, a two-fluid treatment where the gas responds dynamically, generalization to three dimensions, and non-linear growth. Stirring by turbulent fluctuations can also be analyzed in more detail, including the use of direct numerical simulations. 

\acknowledgements 
This work benefitted greatly from helpful suggestions by Jeremy Goodman and Frank Shu.  I thank Bill Ward, Scott Tremaine, and Aristotle Socrates for stimulating discussions.  This material is based upon work supported by the National Aeronautics and Space Administration under Grant NAG5-11664 issued through the Office of Space Science.

\appendix

\section{A. Proof of Stability Criterion}\label{sec:proof}
Proof of the $F < 1$ stability criterion follows.  We express the dispersion relation as \begin{equation}
P(\gamma) \equiv \gamma^3 + {2 } \gamma^2/\tau_s + (\tau_s^{-2} + F)\gamma + {(F-1)/ \tau_s}= 0\, . 
\end{equation}
Properties of the real roots follow readily from Descartes' ``rule of signs"  (J. Goodman, personal communication).  With $F < 1$ there is one sign change to the coeffients of $P(\gamma)$ and thus one and only one purely growing mode.  For $F>1$ there are no purely growing modes, and the coefficients of $P(-\gamma)$ show that there are 1 or 3 purely damped modes.

To complete the proof we show that there can be no complex roots with a positive real part, which is true for all $F$.  Suppose that $P$ has one real root, $\gamma_1$, and a pair of complex conjugate roots, $\gamma_2$ and $\gamma_3 = \gamma_2^*$.  From $(\gamma - \gamma_1)(\gamma - \gamma_2)(\gamma - \gamma_2^*) = P(\gamma)$,  the quadratic term requires that the real part of the complex roots $\Re(\gamma_2) = \Re(\gamma_3) = -\gamma_1/2 - \tau_s^{-1}$.   We (fallaciously) suppose that $\Re(\gamma_2) > 0$.  Then $\gamma_1 < -2/\tau_s$, requiring $F > 1$ and $P(-2/\tau_s) > 0$ (with only one real root, $P > 0$ for all $\gamma > \gamma_1$).  Since $P(-2/\tau_s) = -2/\tau_s^3-(1+F)/\tau_s < 0$, the contradiction proves that there are no overstable oscillations (growing complex roots).  Thus $F<1$ is a necessary and sufficient condition for instability, and growing modes have no oscillatory component.

\begin{figure}[tb]
   \centering
   \includegraphics[width=5in]{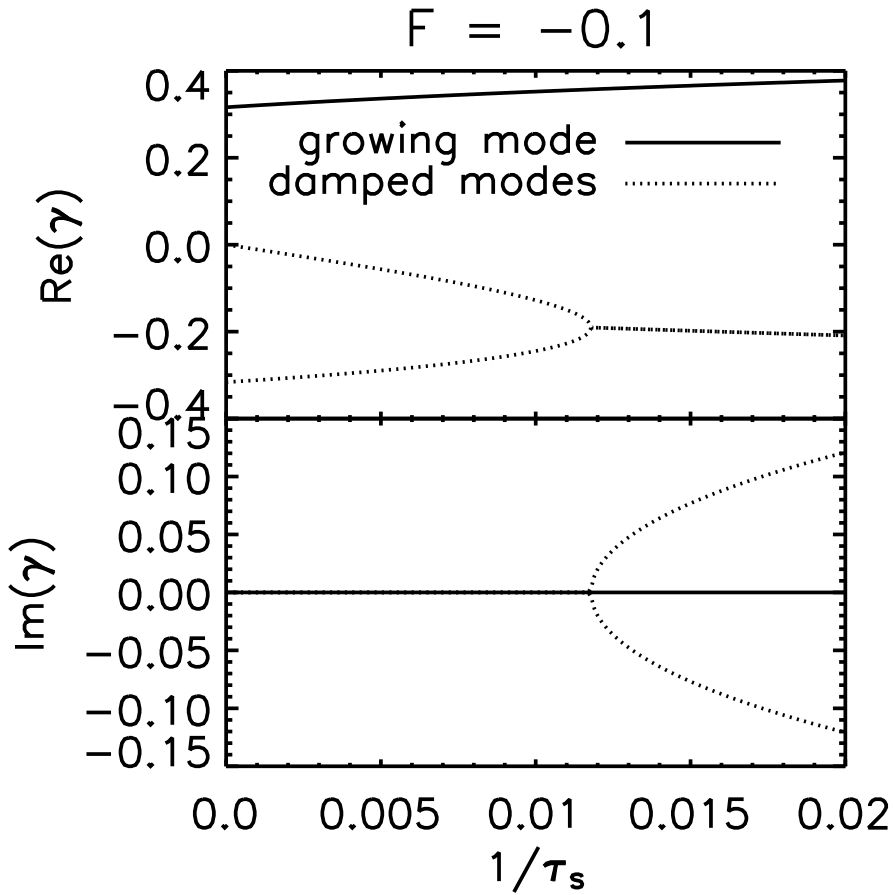} 
  \caption{Growth rates (\emph{top}) and oscillation frequencies (\emph{bottom}) of growing (solid curve) and damped (dotted curves) modes in the loose coupling regime.  With $F = -0.1 < 0$, self-gravity is strong enough to produce GI without dissipation.  Bifurcation corresponds to the changing character of modes.  See \S\ref{weak}.}
   \label{fig:bifurcate}
\end{figure} 

\section{B. Loose Coupling}\label{weak}
The loose coupling limit may be relevant to planetesimal formation in the outer disk or for the subsequent assemblage of larger bodies like planetesimals or planetesimal fragments.  It also establishes the link between destabilized neutral modes and unstable density waves.  If $t_{\rm grow} = 1/(\gamma \Omega) < \ts$, i.e.\  $\gamma \tau_s > 1$, the fluid approximation is poor and a kinetic theory treatment would be more appropriate.  When $\tau_s \gg 1$ approximate formulae for  the growth rate fall in one of three regimes:
\begin{subnumcases} {\label{tsbig} \gamma \approx}
\sqrt{-F} + {1 + F \over -2 F \tau_s}  & \text{if $F \lesssim -\tau_s^{-2/3}$,}\label{tsbiga}\\
 {1 - F \over  F \tau_s } & \text{if $\tau_s^{-2/3} \lesssim F$,} \label{tsbigb} \\
\tau_s^{-1/3}-{F} \tau_s^{1/3}/3 &\text{if $|F| \lesssim \tau_s^{-2/3}$.}\label{tsbigc}
\end{subnumcases}
The first case, with $F < 0$ but not too close to $0$, is an unstable density wave  with a small drag correction that increases the growth rate.  This is seen in Figure \ref{fig:cubgrow} where  contours are nearly horizontal for $F < 0$,  $\tau_s \gg 1$.  For case two, $F > 0$ by a finite amount, and the neutral mode is destabilized for $F < 1$.  Case three shows that the transition across $F = 0$ is well-behaved.  

This transition is associated with a bifurcation point at $F \simeq - 0.52 \tau_s^{-2/3}$, see Figure \ref{fig:bifurcate}.  To the left of the bifurcation point, as $\tau_s \rightarrow \infty$, the unstable mode is an ordinary density wave.   To the right of the bifurcation point, the unstable mode is characterized as a neutral mode, and for $\tau_s \ll 1$ connects with  equation (\ref{drtight}).   Note that the bifurcation actually involves the damped modes. When $F > 0$, there are no unstable density waves, or bifurcation points, and the unstable mode can always be identified as a neutral mode.

Growth rates of the fastest growing modes in the $F > 0$ regime are approximated  by
\begin{subnumcases}{\label{looselong}\gfgm  \approx}
 {1 \over \tau_s Q_T^2} & if $\kfgm h \ll 1$,\\
{2 \over \tau_s( Q_R - 2)} & if $\kfgm h \gg 1$,
\end{subnumcases}
where $Q_R > 2$.   As in the tight coupling case,  growth in the long (or short) wave regime depends primarily on $Q_T$ (or $Q_R$), respectively.

\section{C. Velocity Dispersion: Ignored Effects}\label{sec:dc}
\subsection{Aerodynamic Drift}
Since we have used a single particle size,  aerodynamic drift is uniform and has no effect on random speeds.  
Generalization to a dispersion of particle sizes (or more precisely, stopping times) would give a range of drift speeds, with the radial component being relevant for axisymmetric collapse.  We now show that drift speeds are smaller than velocities induced by turbulent forcing, $c$, except perhaps near marginal coupling.   

The inward radial drift speed of solids is (\citealp{nsh86}, YG05):
\begin{equation}\label{vdr}
\vdr = 2 \mu_g^2  {\eta v_K \tau_s \over 1 + \mu_g^2 \tau_s^2}
\end{equation}
where $\mu_g \equiv \rho_g / (\rho + \rho_g)$ is the midplane gas fraction and $\eta \sim c_g^2/v_K^2$ measures radial pressure support.  We ignore the effect of turbulent diffusion on radial flow \citep{tak02}.  For $\tau_s \ll 1$ and $c \approx \sqrt{\ag} c_g$ (\S\ref{orbturb}) a bit of manipulation shows:
\begin{equation}
{c \over \vdr} \sim {\sqrt{\ag}  } {v_K /c_g\over  \mu_g^2  \tau_s} \sim {h \over h_g}{v_K /c_g \over \mu_g^2  \sqrt{\tau_s}} \sim {\Sigma \over \Sigma_g} {v_K /c_g \over \mu_g \mu_p \sqrt{\tau_s}} \gg 1
\end{equation}
Where we used equation (\ref{hkep}) and $\mu_p = \rho/\rho_{\rm tot} = 1-\mu_g$ is the solid mass fraction.    The inequality holds because even with $\Sigma/\Sigma_g \sim .01$ in the final expression, the contributions to the second fraction  are all $> 1$ and will dominate.   Surprisingly, turbulent speeds dominate even for very weak turbulence.  This is because the particle layer becomes so thin and dense that  $\vdr$ decreases even faster due to the $\mu_g^2$ term.  For $\tau_s \gg 1$, we can similarly show that radial drift is ignorable, but drift could contribute to the velocity dispersion of marginally coupled solids.

\subsection{Collisional vs.\ Aerodynamic Dissipation}
We have ignored the role of inelastic collisions on the velocity dispersion, an effect that could make the disk more gravitationally unstable.   The relative importance of collisions and drag depends on the applicable drag law \citep{ahn76, stu77}, the relative abundances of solids and gas, and the ratio $c/(\Omega h)$.  In the Epstein regime, the ratio of drag to collisional timescales,
\begin{equation}
{\ts^{\rm Ep} \over t_{\rm coll}} \sim {\rho_s a \over \rho_g c_g}{\Sigma c \over h \rho_s a}  \sim {\Sigma \over \Sigma_g} {c \over \Omega h} \ll 1,
\end{equation}
is independent of particle size, $a$, and drag dominates when gas is more abundant than solids ($\Sigma_g > \Sigma$) and/or $c \ll \Omega h$ as found in \S\ref{orbturb}.  If turnover times are short (\S\ref{sonicturb}) and $c \gg \Omega h$ then inelastic collisions would provide additional dissipation.

In the Stokes regime, particles are bigger than the gas mean free path, $a > \lambda_{\rm mfp}$, but not large enough to trigger turbulent wakes.  In this case 
collisions are relatively more important:
\begin{equation}
{\ts^{\rm St} \over t_{\rm coll}} \sim {\Sigma \over \Sigma_g} {c \over \Omega h} {4 a \over 9 \lambda_{\rm mfp}}.
\end{equation}
However increasing $a$ likely brings particles to the turbulent drag regime before collisions dominate.  For large particles in the turbulent drag regime:
\begin{equation}
{\ts^{\rm turb} \over t_{\rm coll}} \sim {8 \over 3 C_D} {\Sigma \over \Sigma_g} {c \over \Omega h} {c_g \over \Delta v} \approx 6 {\Sigma \over \Sigma_g} {v_K \over c_g}\, .
\end{equation}
In the second equality we use the drag coefficient $C_D = 0.44$, and since $\tau_s \gg 1$ for turbulent drag,\footnote{However, high gas densities and short orbital periods, could give $\tau_s < 1$  in the turbulent drag regime.} we take $c \approx \Omega h$  and the relative velocity $\Delta v = \eta v_K \approx c_g^2/v_K$.  Thus for turbulent drag, collisions and drag have comparable importance at cosmic abundances $\Sigma_g \approx 100 \Sigma$, and collisions are more significant when gas is depleted.  

Our dynamical model uses a linear drag law, either Epstein's or Stokes'.  In these regimes collisional dissipation is significant if turbulent stirring is vigorous and $c \gg \Omega h$.  Thus solids may be more gravitationally unstable than our model indicates.  Of course when particle overdensities are highly non-linear (a stage we do not describe in this work) collisional damping will dominate drag.

\bibliography{refs}
\end{document}

%% file: tab1.tex
\begin{deluxetable}{llll}
\tablewidth{0pt}
\tablecaption{Symbols \label{tab:symb}}
\tablehead{\colhead{Symbol} & \colhead{Reference Eq.} & \colhead{Meaning}}

\startdata
$\Omega$ &  & orbital frequency\\
$\ts$, $\tau_s$ & eq.\ (\ref{ts})
& particle stopping time \\
$\Sigma$ & & particle surface density\\
$c$, $h$ & eqs. (\ref{ctot},\ref{hgen}) & particle random speed, scale height\\
$Q_T$, $Q_R$ & eqs.\ (\ref{QT}, \ref{QR}) & stability parameters for solids\\
$\rho$ & $\Sigma/h$ & particle space density\\
$c_g$, $h_g$& $h_g = c_g/\Omega$ &  gas sound speed, scale height\\
$\alpha_g$&eq. (\ref{eq:nu}) &turbulent diffusion parameter \\
$t_0$,$v_0$,$l_0$& $\tau_0 = \Omega t_0$ &eddy turnover time, speed, size\\
$\gamma$ & eq.\ (\ref{eq:gamma}) & dimensionless growth rate \\
$k$ ($\kfgm$)& &wavenumber (fastest growing)\\
$\mathcal{K}$ ($\Kfgm$) & eq.\ (\ref{eq:K})& dimensionless wavenumber \\
$\lambda_{G}$ & $2\pi^2G\Sigma/\Omega^2$ 
& traditional GI wavelength\\
$R$& & cylindrical disk radius \\
$x$, $y$ & & local radial, azimuthal coords. \\
$Q_{\rm sonic}$& eq.\ (\ref{Qsonic})
& ``mixed" Toomre parameter\\
\enddata

\end{deluxetable} 